\documentclass[aps,prl,twocolumn]{revtex4-2}
\usepackage{amsmath,amssymb,amsfonts}
\usepackage{bm, latexsym}
\usepackage{graphicx}
\usepackage{bbm,times}
\usepackage[normalem]{ulem}
\usepackage[usenames,dvipsnames]{color}
\usepackage[pdftex]{hyperref}
\usepackage{braket}
\usepackage{mathtools}

\begin{document}

\newcommand{\cred}[1]{\textcolor{black}{#1}}
\newcommand{\ccred}[1]{\textcolor{black}{#1}}


\title{State-Selective Floquet Memory in Degenerate Manifolds}

\author{Ayan Sahoo}
\affiliation{Harish-Chandra Research Institute, Chhatnag Road, Jhunsi, Prayagraj 211 019, India\\
Homi Bhabha National Institute, Training School Complex, Anushakti Nagar, Mumbai
400 094, India}
\author{Argha Debnath}
\affiliation{Harish-Chandra Research Institute, Chhatnag Road, Jhunsi, Prayagraj 211 019, India\\
Homi Bhabha National Institute, Training School Complex, Anushakti Nagar, Mumbai
400 094, India}
\author{Debraj Rakshit}
\email{debraj.rakshit@hri.res.in}
\affiliation{Harish-Chandra Research Institute, Chhatnag Road, Jhunsi, Prayagraj 211 019, India\\
Homi Bhabha National Institute, Training School Complex, Anushakti Nagar, Mumbai 400 094, India}

\date{\today}

\begin{abstract}
In order to understand which initial states retain memory under periodic driving, it is necessary to depart from the ideal limit giving rise to invariant structures and work inside near-degenerate manifolds. We show that degeneracy itself is not conclusive and what actually matters is how perturbations, such as structural and driving imperfections, acts on the projected multiplet. For systems decoupling into the local clusters in the ideal limit, the projections can be computed within degenerate perturbation theory. There the stability is decided by two criteria: the coupling can not connect the state with other degenerate partners in the manifold, either by kinetic blocking, or more generally, by diagonalizing the projected coupling, and the drive's local selection rule must admit only transfer energies detuned from Floquet sidebands, in which case the projected drive generator exactly vanishes. We formalize these understandings in a periodically driven clean, short-ranged alternating XXZ chain  and demonstrate a set of results that include, long-lived states, both of product form and entangled across the clusters, a periodic family of sideband resonances, and an explicit example showing opposite fates for states with identical energy and charges.
\end{abstract}

\maketitle

\textit{Introduction}--Routes for preventing heating in the periodically driven interacting systems include disorder- and static field-gradient induced localization \cite{Khemani2016,Else2016,von2016, von2016b,Liu2023,sahoo2025power,Li2012, Wilczek2013,Choi2017,Zhang2017,Wang2025}, long-range interactions \cite{Bhakuni2021,Ho2018, Machado2019, Machado2020, Kuwahara2016}, sufficiently high driving frequencies \cite{Abanin2015, Mori2016, Abanin2017a, Abanin2017b, Weidinger2017, Singh2019, Santos2021,Pizzi2021, McRoberts2023, Dutta2025, ghosh2026floquet, Rubio2020,Ghosh2023}, and kinetic constraints or Hilbert-space fragmentation \cite{Sala2020, Khemani2020may, Scherg2021, Kohlert2023, Adler2024, Will2024, Honda2025, Zhao2025}. But, which particular states are protected? In localized phases without mobility edge, the protection is not state-selective and the suppression of relaxation happens through emergent conservation laws distinguishing initial product states \cite{Serbyn2013,Lazarides2015,Ponte2015}. In the constrained models
, such as in the scarred, fragmented, and dynamically frozen systems, protection mechanism is associated with particular invariant structures \cite{Shiraishi2017Systematic,Moudgalya2022Scars,Haldar2021Dynamical, Das2010,Lu2026}. 

Protection by invariant structures arises in an ideal limit, whereas a stable physical system must survive  structural and dynamical perturbations. The decoupled local structure gives rise to extensive conserved quantities and exponentially many invariant sectors. The local degeneracies arrange the states into multiplets. Realistic perturbations generally destroy the local constraints. What remains is an approximate block structure. Structural perturbation can cause loss of fidelity by rotating it within the degenerate manifold. Quasienergy absorption occurs via driving imperfection, which can destroy the memory irreversibly. Enhanced symmetry  and quasienergy folding produce huge degeneracies, and indeed, there, no perturbation is small: chaos is known to set in at arbitrarily weak coupling in finite size \cite{Abdelshafy2025O}.

Degeneracy neither destroys nor protects a state. What becomes crucial is the action of the projected perturbations.  The projections can be computed at the ideal limit, where the system is decomposable as local clusters. Whether the structural perturbation mixes states within a multiplet is determined by the allowed local moves, allowing connection with its degenerate partners.  Blocking them leaves the projected coupling without off-diagonal elements. Whether the drive imperfection can absorb quasienergy is decided by a set of local excitation energies. When none of them matches a Floquet sideband, the projected generator vanishes in the manifold. Neither condition is automatic: any rearrangement directly enabled by the coupling, if activated  at first order, and a generator with resonant transfer energies,  is immediately detrimental. 


We theorize these understandings in a minimal setting--a periodically kicked alternating XXZ chain, where the clusters are strong-bond dimers and the emergent charge counts polarized dimers. There all the ingredients that hosts multiplet structure, kinetic blocking, emergent selection rules, and the resonance conditions. We demonstrate predicted long-lived states, both of product form and entangled across the clusters, sideband resonances and a clear evidence showing that two states with identical energy and charges can have opposite fates, via an explicit example. 

\textit{Systems and driving protocol}--We consider an alternating XXZ chain described by the Hamiltonian $ H(\lambda) = H_0+\lambda H_1$, where
\begin{align}
    H_0 = \sum_{i=1}^{L/2} h_{2i-1,2i}, \quad
    H_1 = \sum_{i=1}^{L/2} h_{2i,2i+1}.
\end{align}
Here, \(H_0\) contains the strong bonds, whereas \(\lambda H_1\)
contains the weak bonds. The two-site XXZ interaction is
\begin{align}
h_{j,j+1} &= J \left( \sigma^x_j\sigma^x_{j+1}
+ \sigma^y_j\sigma^y_{j+1}
+ \Delta\, \sigma^z_j\sigma^z_{j+1} \right),
 \label{eq:Hamiltonian}
\end{align}
where $\lambda$ controls the degree of dimerization between strong and weak bonds and $\Delta$ is the XXZ anisotropy parameter. 

The chain is subjected every $\tau$ to an imperfect periodic global $\pi$ pulse about the $x$ axis, giving the Floquet operator
\begin{align}
U_F(\lambda,\epsilon)
 &=e^{-i(1-\epsilon)G}e^{-i\tau H(\lambda)}
   =X e^{i\epsilon G}e^{-i\tau H(\lambda)},
\label{eq:floquet}
\end{align}
where $\tau$ is time-period, $G=\frac{\pi}{2}\sum_{j=1}^{L}\sigma_j^x$, with $X=e^{-iG}=(-i)^L P_x$ and
$P_x=\prod_j\sigma_j^x$. Since every XXZ bond is invariant under the
global spin flip, $[H(\lambda),P_x]=0$; thus, at $\epsilon=0$, the kick
is, up to a global phase, precisely $P_x$. The stroboscopic evolution
is $|\psi(n\tau)\rangle=U_F^n|\psi(0)\rangle$.

\textit{Dimerization and polarized-dimer number}-- For $\lambda \ne 1$, the system is featured with alternate strong and weak bonds \cite{Marquez2024,Yamanaka1993}. The ideal limit structure can be explicitly followed from a single strong bond connecting two adjacent sites, say $(2i-1,2i)$. The two-spin eigenstates of the local Hamiltonian on this bond consist of one singlet state $|s \rangle_{2i} = \frac{1}{\sqrt{2}}\left(|\uparrow \downarrow \rangle - |\downarrow\uparrow\rangle\right)_{2i}$, with energy $-(\Delta+2)$ and three states corresponding to the usual triplet sector--two polarized states, $|t_u\rangle_{2i}=|\uparrow \uparrow\rangle_{2i}$ and $|t_d\rangle_{2i}=|\downarrow \downarrow\rangle_{2i}$, both with energy $\Delta$, and a symmetric zero-magnetization state, $|0\rangle_{2i}=\frac{1}{\sqrt{2}}\left(|\uparrow\downarrow\rangle+|\downarrow\uparrow\rangle\right)$,
with energy $(-\Delta+2)$.

On each strong bond we define the polarized-dimer projector
\begin{align}
p_i=\frac{1+\sigma_{2i-1}^z\sigma_{2i}^z}{2},
\qquad
P_{\rm dimer}=\sum_{i=1}^{L/2}p_i ,
\label{Pdimer}
\end{align}
where $p_i=1$ on $|t_u\rangle_i$ and $|t_d\rangle_i$, and $p_i=0$ on $|0\rangle_i$ and $|s\rangle_i$. In the ideal decoupled limit, $[H_0,p_i]=[X,p_i]=0$, and hence $[U_F(0,0),p_i]=0$, $\forall i$; consequently, $P_{\rm dimer}$ is exactly conserved in the decoupled limit, and so is the dimer parity, defined as $\Pi=(-1)^{P_{\rm dimer}}$.


\textit{Numerical evidence of stable long-lived states}--We compute the stroboscopic fidelity,  $\mathcal{F}_n=|\langle\psi(0)|U_F^n|\psi(0)\rangle|^2$, at even $n$. Figure~1 shows $\mathcal{F}_n$ for the initial state $|\uparrow\uparrow\cdots\uparrow\rangle$. The fidelity is near-unity over a plateau extending to $10^{8}$--$10^{9}$ driving cycles, after which a loss of fidelity is observed. The extracted lifetimes grow exponentially with system size (inset).  Throughout, $t^{*}$ denotes the first crossing of the fidelity threshold, chosen as $\mathcal{F}_n\geq0.9$.  Further numerical investigations show formation of the fidelity plateau for  the singlet product $\otimes_i|s\rangle_i$, and low-lying eigenstates of the strongly dimerized regime, among few specific others, whereas generic initial states heat quickly. The remainder of the paper answers the question: why does a plateau form only for certain initial states?

\begin{figure}[t]
    \centering
    \includegraphics[scale = 0.38]{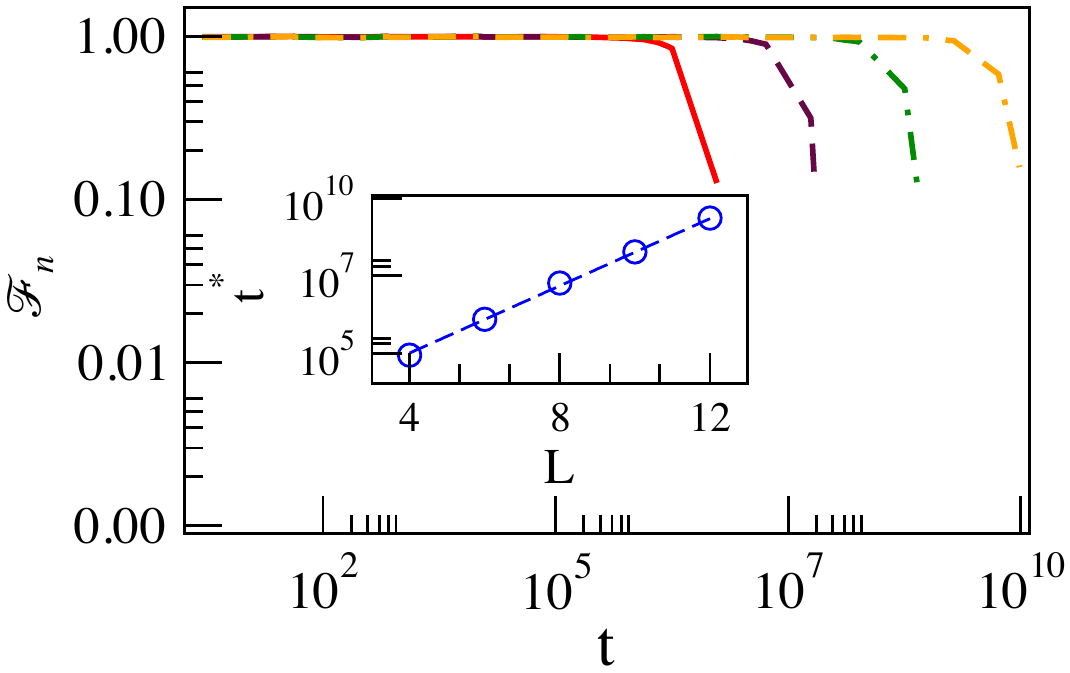}
    \caption{ shows the fidelity $\mathcal{F}_n$ as a function of time $t$ for fixed parameters  $\lambda = 0.01$, $\Delta = 0.1$, $\omega = 4$, and $\epsilon = 0.02$, corresponding to  the initial state  $|\uparrow\uparrow\uparrow\dots\rangle$ for varied system sizes, $L=6$ (red solid), $L=8$ (maroon dash-dash), $L=10$ (green dash-dot) and $L = 12$ (orange dash-dash-dot). Long-lived memory survies till threshold, $t^*$. The inset depicts the numerically extracted $t^{*}$ (circles) as a function of the system size, $L$, and the fitting  (dotted line) from the finite size analysis suggests exponential growth of $t^*$, $t^{*} \sim e^{1.24L}$. }
    
    \label{fig:Fvst}
\end{figure}

\textit{Configurational degeneracy and stroboscopic synchronization—}In the limit $\lambda=0$, the strong bonds are mutually decoupled. The  degeneracy structure of the Floquet operator in this limit can be easily followed by working in the joint eigenbasis of $h_{2i-1,2i}$ and the corresponding two-site spin-flip operator $\sigma^x_{2i-1}\sigma^x_{2i}$, with $
|\kappa\rangle\in \{|+\rangle,|-\rangle,|0\rangle,|s\rangle\}$, where $|\pm\rangle={|t_u\rangle\pm|t_d\rangle}/{\sqrt{2}}$.
At $\lambda=\epsilon=0$, the Floquet eigenstates are the dimer-product states: $H_0|\alpha\rangle=E_\alpha|\alpha\rangle$, where $
|\alpha\rangle=\bigotimes_{i=1}^{N}|\kappa_i\rangle$ and  $N=L/2$. Their Floquet eigenphases are determined by
\begin{equation}
U_F(0,0)|\alpha\rangle=e^{-i\phi_\alpha}|\alpha\rangle,
\qquad
\phi_\alpha=\bigl(\tau E_\alpha+\pi k_\alpha\bigr)\bmod 2\pi,
\end{equation}
where $\chi_\alpha=(-1)^{k_\alpha}$. 
Both the perturbations, $\lambda$ and $\epsilon$, preserve $\chi$, since $[H_1,P_x]=[G,P_x]=0$. Therefore, within a fixed-$\chi$ sector, two states share the same Floquet eigenphase when
\begin{equation}
\Delta E_{\beta\alpha}=r\omega,
\quad r\in\mathbb{Z},
\label{Floquet-degen-condition}
\end{equation}
where $\Delta E_{\beta\alpha}=E_\beta-E_\alpha$ and $\omega=2\pi/\tau$. We define the corresponding eigenphase multiplets and their projectors as
\begin{equation}
    \mathcal{M}_{\phi}=
    \text{span}
    \bigl\{ |\alpha\rangle:\phi_{\alpha}=
    \phi \bigr\}, \qquad
    P_{\phi}=\sum_{\alpha:\phi_{\alpha}=\phi} |\alpha\rangle\langle\alpha|.
\end{equation}
Configurations with same occupation multiset $(N_+,N_-,N_0,N_s)$ possess identical $E_\alpha$, $k_\alpha$, and $\phi_\alpha$. More generally, true-energy degeneracies ($r=0$) and stroboscopic synchronization ($r\neq0$) coexist inside the same eigenphase multiplet.


\textit{Two perturbative channels}--Expanding the Floquet operator to first order in the two perturbations using the standard parameter-derivative identity for operator exponentials \cite{Shirley1965,RodriguezVega2018,Eckardt2015,Blanes2009,Kuwahara2016,Wilcox1967}
\begin{equation}
U_F(\lambda,\epsilon)=U_F^{0}+\lambda {\mathcal V}_\lambda+\epsilon {\mathcal V}_\epsilon
+O(\lambda^2,\epsilon^2,\lambda\epsilon),
\label{Eq:first-order-expansion}
\end{equation}
where
\begin{equation}
\mathcal{V}_\lambda=-i U_F^{0} \int_0^\tau dt
e^{iH_0t}H_1e^{-iH_0t},
\qquad
\mathcal{V}_\epsilon=i G U_F^{0}.
\label{Eq:interaction-picture}
\end{equation}
The corresponding matrix elements are
\begin{equation}
\langle\beta|\mathcal{V}_\lambda|\alpha\rangle
=-e^{-i\phi_\beta}(H_1)_{\beta\alpha}
f(\Delta E_{\beta\alpha}),
\label{mat_elem_Vlamda}
\end{equation}
and
\begin{equation}
\langle\beta|\mathcal{V}_\epsilon|\alpha\rangle
= i e^{-i\phi_\alpha} G_{\beta\alpha},
\end{equation}
where $f(\Delta E)=(e^{i\tau\Delta E}-1)/\Delta E$. For stroboscopically synchronized states with $\Delta E=r\omega$, $r\neq0$, the period-average factor in the $\lambda$ channel vanishes. Hence its secular action is restricted to true-energy-degenerate sectors and we work inside the projected manifold \cite{Schrieffer1966,Bravyi2011,MacDonald1988}. In terms of the projector $P_{E}$ onto states of energy $E$ inside $\mathcal M_\phi$, its secular matrix reduces to 
\begin{equation}
\mathcal{V}_\lambda^\phi=-i\tau e^{-i\phi}
\sum_{E} P_{E} H_1 P_{E},
\label{lambda-matrix-elm}
\end{equation}
where $P_{E}=\sum_{\alpha:E_\alpha=E} |\alpha\rangle \langle \alpha |$. The kick is instantaneous without involving period-averaging. Consequently, 
\begin{equation}
\mathcal{V}_{\epsilon}^{\phi} = i e^{-i\phi} P_{\phi} G P_{\phi},
\label{kick-matrix-elm}
\end{equation}
and hence, can therefore hybridize  stroboscopically synchronized states. These two channels identify three routes to fidelity loss: secular rotation within a true-energy-degenerate multiplet through the $\lambda$ channel, irreversible quasienergy absorption through the $\epsilon$ channel, and, at Floquet sideband resonance, an $O(1)$ kick-driven rotation within a common eigenphase multiplet.


\textit{Stability in $\lambda$-channel: Protection criteria}--Fidelity loss can happen without violating dimer parity through ungapped secular dynamics inside a degenerate multiplet. Let $\mathcal{M}_E$ be the degenerate eigenspace of $H_0$ containing $|\psi\rangle$, i.e.,   only the  $\Delta E=0$, the true-energy-degenerate part of $\mathcal{M}_\phi$ in Eq.~(5), is considered. For an eigenstate of $P_E H_1 P_E$, $\lambda$-channel produces no $O(1)$ secular mixing within the degenerate subspace on the plateau timescale, but only a bounded dressing of order $(\lambda/\delta_{\mathrm{off}})$, where $\delta_{\mathrm{off}}$ is smallest energy cost of an $H_1$-allowed transition out of the degenerate manifold. Allowed dressing obeys the weak-bond selection rule $\Delta P \in \{0,\pm2\}$, preserving the dimer parity $\Pi = (-1)^{P_{\rm dimer}}$ exactly. 

Consider a general dimer product state,
$|\Psi\rangle=\otimes_{k}|\psi_k\rangle$,
$\psi_k\in\{|t_u\rangle,|t_d\rangle,|0\rangle,|s\rangle\}$. The relevant amplitudes is $\mathcal{J}_{\kappa\kappa'} \equiv
\langle \kappa' \kappa |h_{2i,2i+1} |\kappa \kappa' \rangle$, where $\kappa \neq \kappa'$. The matrix elements are $\mathcal{J}_{t_u s}=\mathcal{J}_{t_d s}=-1,
\mathcal{J}_{t_u 0}=\mathcal{J}_{t_d 0}=+1,
\mathcal{J}_{0 s}=-\Delta,
\mathcal{J}_{t_u t_d}=0$.  Any adjacent pair mixing a polarized dimer with an unpolarized one, e.g.,
($t_{u}$ or $t_{d}$  next to $s$ or $t_0$), or an $s$ next to a $t_0$, exchanges
at first order. Its fidelity locally dephases as, $n_{\rm hop}\sim \big(\lambda |\mathcal{J}_{\kappa\kappa'}| \tau\big)^{-1}$, independently of $\epsilon$ and of $P_{\rm dimer}$ conservation. 

\begin{figure}[t]
    \centering
    \includegraphics[scale = 0.5]{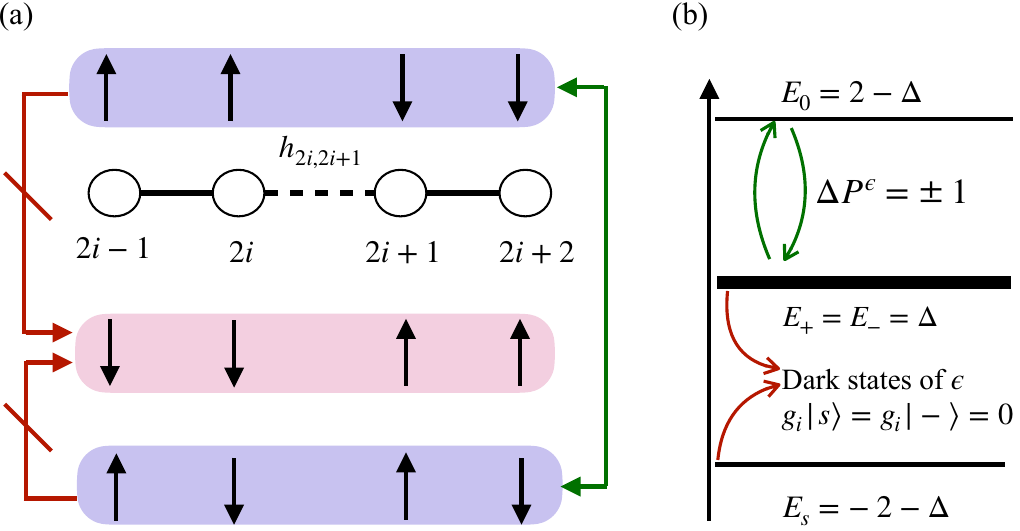}
    \caption{(a) Kintetic blockade in the $\lambda$-channel: The off-diagonal part of the weak-bond interaction, $h_{2i,2i+1}$, flips the inner spins, when they are opposite, while leaving outer spins unchanged. Under any power of $h_{2i,2i+1}$, $|t_ut_d\rangle=|\uparrow\uparrow\downarrow\downarrow\rangle$ can connect to a configuration with flipped inner pair of spins, but can not rearrange to the configuration of its degenerate partner $|t_dt_u\rangle=|\downarrow\downarrow\uparrow\uparrow\rangle$ [Eq.~14]. (b) Selection rules of the $\epsilon$-channel: The local kick $g_i$ generator couples only $|+\rangle$ and $|0\rangle$. In contrast, it annihilates $|-\rangle$ and $|s\rangle$ exactly, making them dark to all orders in $\epsilon$.}
    
    \label{fig:Fvst}
\end{figure}
The vanishing of $\mathcal{J}_{t_u t_d}$ turns out to be the key. The weak bond $h_{2i,2i+1}$ acts only on
the inner sites $(2i,2i+1)$, whereas exchanging
$|t_u\rangle$ and $|t_d \rangle$ on the adjacent dimers requires flipping the outer sites $(2i-1,\,2i+2)$ as well. However, no power of $h_{2i,2i+1}$ acts on the edges. Hence, 
\begin{equation}
\langle \cdots t_d^{(2i)}t_u^{(2i+1)} \cdots| h_{2i,2i+1}|\cdots t_u^{(2i)} t_d^{(2i+1)} \cdots\rangle = 0,
\label{kineticblock}
\end{equation}
is kinetic blockade of the connecting bond that persists even in higher order (see Fig.~2(a)). Rearranging a polarized pattern therefore requires intermediate states outside the polarized sector, and is comparatively far weaker.

\textit{$\lambda$-Protected states}--The maximal charge sector is the protected manifold as a consequence of  kinetic blockade in (\ref{kineticblock}) For the fully polarized manifold, the unperturbed eigenspace
$\mathcal{M}_E$ has energy $E=N\Delta$ and is spanned by the $2^N$ product configurations
$|\alpha\rangle=\bigotimes_i|\alpha_i\rangle$, with
$\alpha_i\in\{t_u,t_d\}$, and $P_E=P_{\rm pol}
=\sum_{\alpha\in\mathcal{M}_E} |\alpha\rangle\langle\alpha| $. On each strong bond, the two polarized states span the local two-dimensional space $\mathcal{H}_i^{\rm pol} = \mathrm{span}\left\{
|t_u\rangle_i,|t_d\rangle_i
\right\}$. We adopt  pseudospin structure \cite{Sachdev1990,Mila1998,Giamarchi2008}, and we define $T_i^z
=|t_u\rangle_i\langle t_u|
- |t_d\rangle_i\langle t_d|$, such that $T_i^z|t_u\rangle_i
= +|t_u\rangle_i$ and $T_i^z|t_d\rangle_i
= -|t_d\rangle_i$. Equivalently, within $\mathcal{H}_i^{\rm pol}$,
$T_i^z=\frac{1}{2}\left(
\sigma_{2i-1}^z+\sigma_{2i}^z
\right)$. This, for periodic boundaries, is manifested in form of the Ising interaction
\begin{equation}
P_EH(\lambda)P_E  =
N\Delta P_E+ \lambda\Delta
\sum_{i=1}^{N} T_i^z T_{i+1}^z.
\end{equation}

We now introduce the pseudospin domain-wall operator
\begin{equation}
\hat{\mathcal{D}}=
\frac{1}{2} \sum_{i=1}^{N}
\left(1- T_i^z
T_{i+1}^z \right).
\label{eq:domain-wall-operator}
\end{equation}
The weak-bond interaction resolves the parent
$2^N$-dimensional eigenspace into first-order multiplets
$\mathcal{M}_{\mathcal{D}}$ having a fixed domain-wall number $D$. Their projectors are
\begin{equation}
P_\mathcal{D}
=
\sum_{\substack{
\alpha\in\mathcal{M}_{E}\\
\mathcal{D}_\alpha=\mathcal{D}
}}
|\alpha\rangle\langle\alpha|,
\qquad
P_E=\sum_\mathcal{D} P_\mathcal{D},
\label{eq:fixed-D-projector}
\end{equation}
where $\dim\mathcal{M}_{\mathcal{D}}=2\binom{N}{\mathcal{D}}$. Within each fixed-$D$ multiplet,
\begin{equation}
P_\mathcal{D}H(\lambda)P_\mathcal{D}|_{\mathcal{M}_\mathcal{D}}=\left(N\Delta+\lambda\Delta(N-2\mathcal{D})\right) {\mathbb I}_\mathcal{D}, 
\end{equation}
where ${\mathbb I}_\mathcal{D}$ is the identity of $\mathcal{M}_\mathcal{D}$, where
$\mathcal{M}_\mathcal{D}=P_\mathcal{D}\mathcal{M}_E$. The uniform polarized states form the $\mathcal{D}=0$ sector, the one-minority configurations constitute a symmetry-resolved subspace of the $\mathcal{D}=2$ sector, and, for even $N$, the two dimer-N\'eel configurations form the extremal $\mathcal{D}=N$ sector. Virtual transitions outside the polarized manifold nevertheless modify the effective Hamiltonian at second order, generating diagonal energy renormalizations and off-diagonal pseudospin exchanges of order $\mathcal{O}(\lambda^2)$. The $\mathcal{D}=0$ ferromagnets are, however, special, as they are exact eigenstates of $H(\lambda)$ for arbitrary $\lambda$.

The mechanisms for remaining protected product states are distinct. The uniform unpolarized products $|\psi^s\rangle$ and $|\psi^0\rangle$ are, for generic $\Delta$, alone in their degenerate subspace, so no secular problem exists and the $\lambda$-channel produces only gapped dressing of order $(\lambda/\delta_{\rm off})$, with $\delta_{\rm off}=O(1)$ set by the two-dimer gaps and is given by
\begin{equation}
    \delta_{\rm off}(\psi)={\rm min}\{\Delta_{\psi \beta}, \beta\notin\mathcal{M}_E, \langle \psi|H_1|\beta\rangle \ne 0\}.
\end{equation}
Noting that $\mathcal{J}_{0 s}=-\Delta$, first-order $\lambda$-freezing occurs trivially for the configurations with $|s\rangle$ next to $|0\rangle$ at $\Delta=0$, i.e., at the XX point.

Finally, we provide direct evidence that no conserved label decides a state's
stability, and that the first criterion is genuinely necessary. For generic
$\Delta$, the one-triplon manifold, one polarized dimer on a singlet
background, is spanned by the $2N = L$ states $|j,\eta\rangle = |s\cdots
t_\eta(j)\cdots s\rangle$, $\eta = u,d$. Every state in it carries the same
local content. The weak bond, in turn, cannot convert one species into the other,
since that would require flipping both spins of the polarized dimer. Up to
a constant, the projected Hamiltonian is therefore a pure hopping problem
on two decoupled lattices,
\begin{equation}
P_E H(\lambda) P_E = \lambda \mathcal{J}_{t_us}\sum_{j,\eta}
\big(|j{+}1,\eta\rangle\langle j,\eta| + \mathrm{H.c.}\big),
\end{equation}
the weak-bond Ising term vanishing on the singlet background. It describes a
triplon of excitation energy $\Delta_t = 2(1+\Delta)$, measured from
$|\psi^s\rangle$. 

$|j,\eta\rangle$, the position eigenstates of the hopping, fails the first criterion and
delocalizes within $n \sim (\lambda\tau)^{-1}$ cycles. Their Bloch superpositions $|k,\eta\rangle = N^{-1/2}\sum_j e^{ikj}|j,\eta\rangle$, with
$k = 2\pi m/N$ and $j$ labelling dimer position, diagonalize it, $E_k = \Delta_t +
2\lambda J_{t_us}\cos k$. These states, $|k\rangle$, satisfy the criterion, and are long-lived. Both cases are picked from the same manifold with identical energy and charge, but have opposite fates. Charge content therefore does not
decide a state's fate, what matters is if the state diagonalizes the
projected coupling \cite{supplementary_comment}.

\textit{Kick selection rules}--The $\epsilon$-channel decides the finite fidelity lifetime of the plateau of the protected states. Expressing $G$ as a sum of the local terms, $G=\sum g_i$, we evaluate their local action on the single dimer spin-basis: $g_i|t_{u}\rangle\propto|0\rangle$,
$g_i|t_d\rangle\propto|0\rangle$,
$g_i|0\rangle\propto|t_u\rangle+|t_d\rangle$, and $g_i|s\rangle=0$. This is presented in Fig.~2(b). Consequently, the kick generator annihilates the singlet product, $G|\psi^s\rangle=0$, and therefore
$e^{i\epsilon G} |\psi^s\rangle=|\psi^s\rangle,~
U_F(\epsilon) |\psi^s\rangle=X |\psi^s\rangle,~ \forall \epsilon$. Hence, the pulse imperfections do not affect the singlet product state, which remains an eigenstate of the system at $\lambda=0$. At finite $\lambda$, however, the non-zero leading amplitude leads to a slow decay with leading-order correction of $O(\lambda\epsilon)$. In the rotated basis: $g_i$ only couples $|+\rangle$ and $|0\rangle$ and annihilates $|-\rangle$ and $|s\rangle$.  However, $\otimes_i|-\rangle_i$ states are not $\lambda$ protected. Thus, different $\lambda$-protected states exhibit different plateau depths. Another implication is that the first-order kick imperfection obeys $\Delta P^{\epsilon}=\pm1$.

At $\epsilon=0$ the Floquet operator $U_F(\lambda,0)$ commutes with the static Hamiltonian and with the dimer parity $\Pi \equiv (-1)^{P_{\rm dimer}}$, i.e., $[U_F(\lambda,0),\Pi]=0,~\forall \lambda$, implying the stroboscopic evolution conserves the energy of $H(\lambda)$ under perfect kick. Driving imperfection is the only source of energy absorption or $P_{\rm dimer}$ violation, causing heating. Off resonance, this heating proceeds through high-order processes, in line with the general picture of slow Floquet heating through high-order absorption \cite{Abanin2015,Mori2016,Mallayya2019Heating} leading to featureless infinite-temperature steady state \cite{Dalessio2014LongTime,Lazarides2014Equilibrium,Ponte2015Periodically}.

\textit{$\epsilon$-channel criteria: Floquet off-resonance}--The protected states' lifetime collapses wherever off-resonant Floquet detuning vanishes, which corresponds to the condition $r\omega=2|1-\Delta|$. The underlying mechanism can be followed from a degenerate Floquet perturbation theory carried out in $\mathcal{M}_\phi$.

The imperfect kick acts non-trivially, in general. Due to the associated selection rules, the local generator $g_i$ annihilates $|s\rangle$ and $|-\rangle$ and couples only $|+\rangle$ and $|0\rangle$. As a result the imperfect kick generator $G$ turns blind on stroboscopic degeneracies for fixed $\chi$ and a nondegenerate perturbation theory can be applied for computing inter-multiplet mixing. For $N$ decoupled dimers, considering two many-body states differing only on the $j^{\rm th}$ dimer, such that the local action of $g_i$ couples $|+\rangle_j$ and $ |0\rangle_j$, the energy difference  between two states connected by one local kick in first order is $\Delta E_{\alpha \beta}=2|(1-\Delta)|$. Hence, inter-multiplet mixing coefficient is then associated with one global Floquet denominator, such that
\begin{equation}
c^{\epsilon}_{\beta\alpha}
=\frac{i\epsilon\, G_{\beta\alpha}}{1-e^{-i\tau\Delta E_{\beta\alpha}}},
\qquad
\big|1-e^{-i\tau\Delta E_{\beta\alpha}}\big|=\mu .
\label{eq:ceps}
\end{equation}
Off resonance, $2|1-\Delta|\neq r\omega$, the projected kick vanishes on the multiplets,  
\begin{equation}
P_\phi G P_\phi=0, 
\end{equation}
implying the first-order eigenphases are unshifted. All admixtures are of $O(\epsilon/\mu)$. Lifetime of the protected states collapses periodically at sideband resonances, $2|1-\Delta| = r\omega$, and at the special isotropic point $\Delta=1$ due to the exact kick degeneracy, rather than a drive induced synchronism \cite{EndMatter}.

\textit{Discussion}--This work formulates criteria essential for protection of the Floqet memory: the coupling can not rearrange the state among its near-degenerate partners and the drive's transfer energies are detuned from Floquet sidebands. We provide a generic prescription for long-lived states that do not require disorder, long-range interactions, high-frequency driving and fine-tuning of the parameters.

The setting can be implemented via ultracold atoms in optical superlattices. A spin-chain can be engineered by a two-component Mott insulator \cite{Duan2003Controlling,Trotzky2008Superexchange,Choi2017Exploring}, and bond alternation can be designed via a period-two superlattice with tunable $\lambda$ \cite{SebbyStrabley2006,Greif2013ShortRange}. The anisotropy $\Delta$ can be controlled through spin-dependent interactions \cite{Jepsen2020SpinTransport,Jepsen2021Transverse}. The global kick is realized via stroboscopic driving \cite{Rubio2020}. Other available platforms are superconducting processors \cite{Mi2022TimeCrystalline,Wang2025_superconducting}, trapped ions \cite{Zhang2017}, and Rydberg arrays \cite{Kyprianidis2021Observation}. Natural extensions include exactly solvable models with exotic geometries and constraints \cite{Shastry1981Exact,Miyahara1999Exact,Majumdar1969NextNearestI,Kennedy1987Hidden,Roy2018Response,Kitaev2006Anyons}, where richer local degeneracy could generate richer selection rules.\\

\begin{acknowledgments}
    A. Sahoo acknowledges support from the Infosys Scholarship for Senior Students at the Harish-Chandra Research Institute. We acknowledge the computational resources provided by the cluster computing facility at the Harish-Chandra Research Institute, India. Generative AI tools, ChatGPT and Claude, were used to refine and condense portions of the manuscript, for cross-checking auxiliary calculations for consistency checks, primarily to save time,  and for literature searches.
\end{acknowledgments}




\bibliographystyle{apsrev4-2}
\bibliography{References}

\begin{thebibliography}{81}
\bibitem [{sup()}]{supplementary_comment}%
  \BibitemOpen
  \href@noop {} {}\bibinfo {note} {At $\epsilon = 0$ the kick interchanges the species, $X|k,u\rangle \propto |k,d\rangle$, so the one-cycle Floquet eigenstates are $|k,\pm\rangle = (|k,u\rangle \pm |k,d\rangle)/\sqrt2$, while either species alone revives at even $n$.}\BibitemShut {Stop}%
\end{thebibliography}

\smallskip
\renewcommand{\theequation}{A\arabic{equation}}
\setcounter{equation}{0}

\onecolumngrid
\appendix

\begin{center}
   \textbf{End Matter}
 \end{center}
 
\twocolumngrid
\emph{Perturbative structure for the $\lambda$ and $\epsilon$ channels}--We work with the eigenbasis in the decoupled limit, $\lambda=\epsilon=0$.  The unperturbed Floquet eigenstates satisfy $U_F^0|\alpha\rangle=e^{-i\phi_\alpha}|\alpha\rangle$. We partition the Hilbert space into degenerate eigenphase multiplets $\mathcal M_\phi$ of Eq.~(7), with projectors $P_\phi$. The first-order expansion is given in Eq.~(8), where $V_\lambda$ and $V_\epsilon$ are given in Eq.~(9) in the main text, following which we reexpress $V_{\lambda}$ as $V_\lambda=-iU_F^0 A_\lambda$. The factor   associated with $A_\lambda$ accounts for a phase, whereas $A_{\lambda}$, a function of $H_1$, is Hermitian. Hence, within one eigenphase multiplet,
$ U_F^0P_\phi=e^{-i\phi}P_\phi$, 
and therefore
\begin{equation}
    P_\phi V_\lambda P_\phi = -ie^{-i\phi}P_\phi A_\lambda P_\phi.
\end{equation}

Since both perturbations preserve the spin-flip eigenvalue $\chi$, the secular problem separates into fixed-$\chi$ blocks of each eigenphase multiplet. We work in fixed $\chi$ sector, and drop the index $\chi$ from subscript for convenience. Within such a block, equality of Floquet eigenphases implies $E-E'=r\omega$ (Eq.~6 in the main text). Resolving it further, $ P_\phi=\sum_E P_{\phi,E}$, where $P_{\phi,E}$ projects onto states of energy $E$ inside
$\mathcal M_\phi$. Then
\begin{align}
    P_\phi A_\lambda P_\phi =\sum_{E,E'} I(\tau)
    P_{\phi,E} H_1 P_{\phi,E'},  \\
    I(\tau)=\int_0^\tau dt e^{i(E-E')t} \nonumber,
\end{align}
where $E \neq E'$ accounts for Floquet folding. Using, $I(\tau)=0$ for $E-E'=r\omega$ with $r\neq0$ and and $I(\tau)=\tau$ for $E=E'$, i.e., for the $r=0$ case, we have
\begin{equation}
    P_\phi A_\lambda P_\phi =\tau\sum_E P_{\phi,E}H_1P_{\phi,E}.
\end{equation}

The $\epsilon$-channel analysis proceeds analogously. From Eq.~(9), we have $V_\epsilon=iGU_F^0$. It is instantaneous without involving period averaging. Projection onto $\mathcal M_\phi$ gives $P_\phi V_\epsilon P_\phi = ie^{-i\phi}P_\phi GP_\phi$.  Both projected perturbations have a common phase $e^{-i\phi}$. Factoring it out reduces to Hermitian secular generator governing the first-order splitting inside $\mathcal M_\phi$,
\begin{equation}
A_\phi = \lambda\tau\sum_E P_{\phi,E}H_1P_{\phi,E}
-\epsilon P_\phi GP_\phi.
\label{unified-1}
\end{equation}

\begin{figure}[t!]
    \centering
    \includegraphics[scale = 0.6]{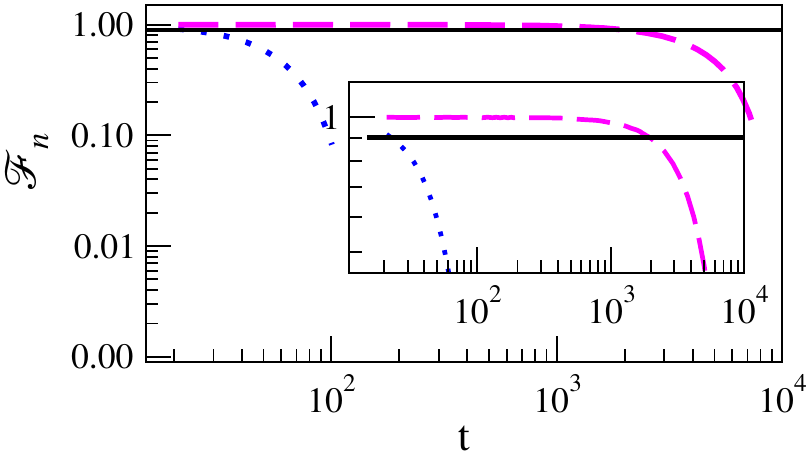}
    \caption{Stroboscopic fidelity
$\mathcal{F}_n$ versus $t=n\tau$ (even $n$) at $\lambda=0.01$,
$\Delta=0.1$, $\epsilon=0.02$, $\omega=3$, $L=8$, for two states of the one-triplon manifold: the localized triplon $|t_u sss\rangle$ (dotted) and the Bloch state
$|k\rangle=\tfrac12\sum_{j=1}^{N}e^{ikj}|j,u\rangle$ and the Bloch state
$|k\rangle=\tfrac12\sum_{j=1}^{N}e^{ikj}|j,u\rangle$, with $|j,u\rangle\equiv|s\cdots t_u(j)\cdots s\rangle$, at $k=0$ (dashed). Both carry the same energy and conserved charges. The solid line shows the threshold, fixed at $\mathcal{F}_n=0.9$.}
\label{F_vs_t_bloch_st}
\end{figure}

The correct zeroth-order states within a degenerate eigenphase multiplet are therefore the eigenvectors of $A_\phi$, $A_\phi|\phi,m\rangle = \delta\phi_m|\phi,m\rangle$, where $m=1,\ldots,\dim\mathcal M_\phi$. Since $A_\phi$ is Hermitian, $\delta\phi_m$ is real. Thus diagonalizing $A_\phi$ fixes the basis ambiguity inside the degenerate multiplet.

The remaining first-order correction to the eigenvector comes from states outside the multiplet. The first-order correction to the eigenvectors involving inter-degenerate subspaces is,
\begin{equation}
|\Phi_{\phi,m}\rangle=|\phi,m\rangle
+\sum_{\beta\notin\mathcal{M}_\phi}
\Big(c^{\lambda}_{\beta m}+c^{\epsilon}_{\beta m}\Big)|\beta\rangle
+O(\lambda^{2},\epsilon^{2},\lambda\epsilon),
\label{unified-2}
\end{equation}
with
\begin{equation}
c^{\lambda}_{\beta m}=\frac{\lambda\,\langle\beta|\delta\mathcal{V}_\lambda|\phi,m\rangle}
{e^{-i\phi}-e^{-i\phi_\beta}},
\quad
c^{\epsilon}_{\beta m}=\frac{\epsilon\,\langle\beta|\delta\mathcal{V}_\epsilon|\phi,m\rangle} {e^{-i\phi}-e^{-i\phi_\beta}}.
\label{coeff-epsilon}
\end{equation}
Within this unified framework, we
evaluate the secular matrices and the the correction coefficients for the two channels.

\emph{$\lambda$-channel}--The period-average result implies that within a fixed-$\chi$ eigenphase multiplet, only the true energy-degenerate blocks contribute to the $\lambda$-channel secular generator. Thus,
\begin{equation}
    A_\phi^{\lambda}
    =\lambda\tau
\sum_E P_{\phi,E}H_1P_{\phi,E}.
\end{equation}
Stroboscopically synchronized states with
$\Delta E=r\omega$, $r\neq0$, play no role in the secular problem. The same cancellation occurs in the inter-multiplet dressing. Consider
a secular eigenvector $|\phi,m\rangle$ belonging to a true-energy block
of energy $E_m$. From Eq.~(10),
\begin{equation}
    \langle\beta|V_\lambda|\phi,m\rangle
=-e^{-i\phi_\beta} (H_1)_{\beta m} \frac{e^{i\tau\Delta E_{\beta m}}-1}
{\Delta E_{\beta m}},
\end{equation}
where $\Delta E_{\beta m}=E_\beta-E_m$. Since the perturbation preserves
$\chi$, we obtain $
e^{-i\phi}-e^{-i\phi_\beta}
= e^{-i\phi_\beta}
\left(e^{i\tau\Delta E_{\beta m}}-1\right)$.
Substitution into Eq.~(A8) therefore gives the exact first-order
cancellation of the Floquet factors,
\begin{equation}
c^\lambda_{\beta m}
=
-\frac{\lambda(H_1)_{\beta m}}
{\Delta E_{\beta m}} .
\end{equation}
The $\lambda$-channel dressing is consequently identical to ordinary static perturbation theory. Noticeably, its denominator contains the true energy
difference rather than a Floquet detuning. 

The $\lambda$-channel protection criterion follows directly from above. If $|\psi\rangle$ is an eigenstate of the projected coupling $P_{\phi,E}H_1P_{\phi,E}$, the $\lambda$ channel produces no first-order secular rotation within the true-energy-degenerate sector. The remaining correction comes only from states outside the multiplet, with amplitudes $c^\lambda_{\beta\psi}$, whose denominators are bounded below by $\delta_{\rm off}$ of Eq.~(19). These admixtures therefore causes only a fidelity deficit of order $O[(\lambda/\delta_{\rm off})^2]$, fixing the plateau depth. Since the denominator in the Eq. (A9) has no Floquet detuning, there is no sideband structure. At higher order, virtual transitions generate effective intra-manifold couplings of order $O(\lambda^2)$, which can lift residual degeneracies and lead to the much slower dephasing. 

\begin{figure}[t]
    \centering
    \includegraphics[scale = 0.4]{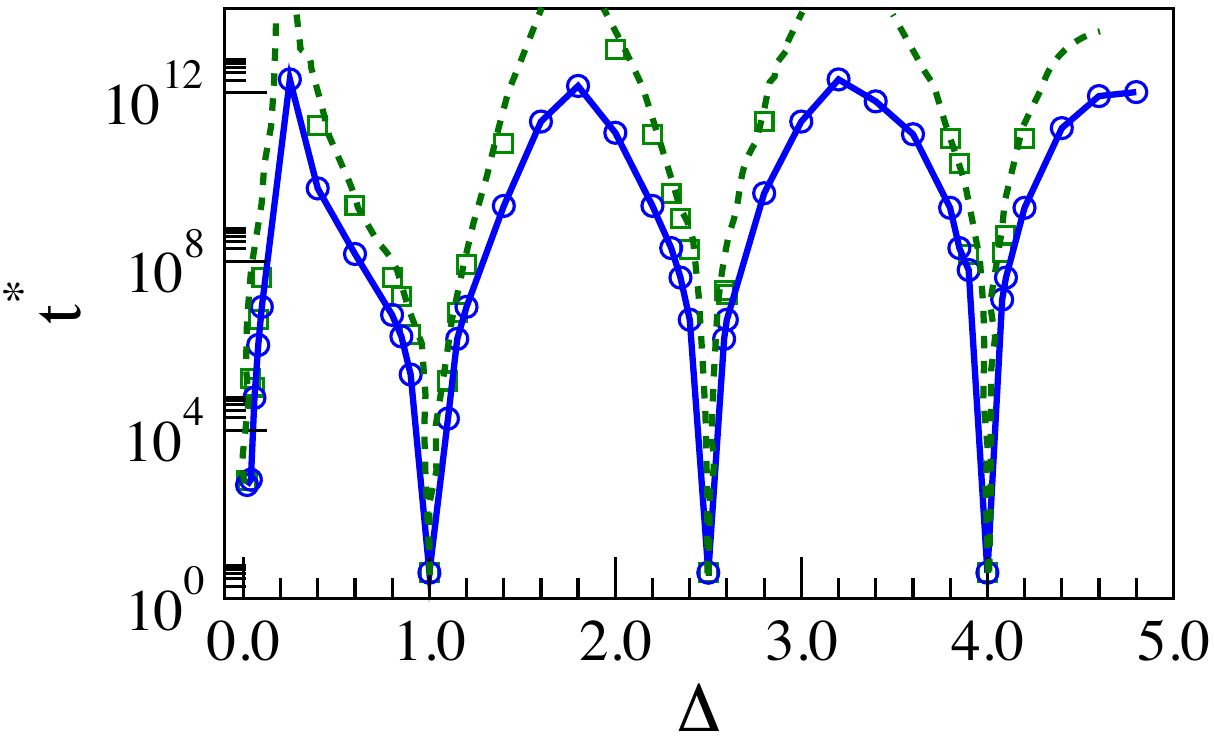}
    \caption{shows $t^*$ as a function of  $\Delta$  corresponding to $\lambda = 0.01$, $\epsilon = 0.02$, and $\omega = 3$ for $L = 10$ (circles), and $12$ (square). Two prominent features can be noticed: off-resonant enhancement of lifetime with system size and periodic collapse due to the Floquet resonance. The lines serve as guide to the eye. }
    \label{lifetime_vs_d}
\end{figure}
Finally, we present a plot showing that while both states belong to the same one-triplon manifold and have the same energy and charge content, the lifetimes are very different. Fig.~\ref{F_vs_t_bloch_st} shows $\mathcal{F}_n$ versus $t$ for $\lambda=0.01$, $\Delta=0.1$, $\epsilon=0.02$, $\omega=3$, and $L=8$ for the state $|t_u sss\rangle$ (dashed) and the Bloch state $|k\rangle=\tfrac12\sum_{j=1}^{N}e^{ikj}|j,u\rangle$, with
$|j,u\rangle\equiv|s\cdots t_u(j)\cdots s\rangle$, at $k=0$ (dashed).  The $|t_u sss\rangle$ state decays rapidly due to first-order mixing through the $\lambda$-channel. On the other hand, the superposition state is long-lived because it is an eigenstate of the projected coupling within the one-triplon manifold. These numerical results confirm that the stability of a state does not depend solely on the conserved charge, but rather on the action of the projected perturbation within the degenerate manifold. 

\textit{$\epsilon$-channel}--
We determine whether the kick imperfection can mix states within an unperturbed eigenphase multiplet $\mathcal{M}_{\phi}$.
The corresponding matrix element due to the kick imperfection is provided in Eq.~({\ref{kick-matrix-elm}}). Unlike the $\lambda$-channel matrix element in Eq.~(\ref{lambda-matrix-elm}), this expression contains no period-average factor
$f(\Delta E)$. The kick acts instantaneously and can therefore access all Floquet sidebands. Whether it produces a secular transition is determined by combining its
local selection rule with the quasienergy-matching condition.

The selection rule states that $G_{\beta\alpha}\neq 0$ only when $|\alpha\rangle$ and
$|\beta\rangle$ differ by
$|\Delta P_{\rm dimer}|=1$, connecting
$|+\rangle_i$ with $|0\rangle_i$ (see 'Kick selection rules' in the main text). Every such pair
satisfies
\begin{equation}
\Delta E_{\beta\alpha}
=
\pm 2(1-\Delta),
\label{G_selection_rule}
\end{equation}
as $k_{\beta}=k_{\alpha}$.
For the same pair to belong to a common eigenphase multiplet, Eq.~(\ref{Floquet-degen-condition}) additionally requires
$\Delta E_{\beta\alpha}=r\omega$. The two conditions can therefore be satisfied only at $2|1-\Delta|=r\omega$, where $r=0,1,2,\ldots$. The case $r=0$ corresponds to the isotropic point
$\Delta=1$. Consequently, away from these resonances, no state connected
by $G$ lies in the same eigenphase multiplet, implying
\begin{equation}
\mathcal{V}_{\epsilon}^{\phi} = i e^{-i\phi} P_{\phi} G P_{\phi}=0.
\end{equation}
Thus, the $\epsilon$ channel produces no first-order secular
rotation within $\mathcal{M}_{\phi}$. 

As, every $G$-connected partner lies in different multiplets, Eq.~(\ref{unified-1}-\ref{unified-2}) applies with a denominator bounded by a non-zero value. Since all $G$-connected pairs share the single energy scale (\ref{G_selection_rule}), the entire first-order correction carries one universal denominator,
\begin{equation}
|\Phi^{(\epsilon)}_{\phi,a}\rangle=|\phi,a\rangle
+ i\epsilon
\sum_{\beta:G_{\beta\alpha}\neq 0}
\frac{G_{\beta\alpha}}
{1-e^{\mp 2i\tau(1-\Delta)}}|\beta\rangle
+O(\epsilon^2),
\label{eq:epsmixing}
\end{equation}
where $\big|1-e^{\mp 2i\tau(1-\Delta)}\big|
=2\Big|\sin\frac{2\pi(1-\Delta)}{\omega}\Big|$. Off resonance the denominator magnitude is uniformly bounded below, the admixture of $\Delta P_{\rm dimer}=\pm 1$ states is $O(\epsilon)$, and quasi-conservation of $P_{\rm dimer}$ holds on every state of the energy-degenerate multiplets.

The lifetime collapses at sideband resonances at
\begin{equation}
2|1-\Delta| = r\omega,
\label{eq:rescond}
\end{equation}
the $G$-connected states merge into common multiplets, the secular problem has a nonzero the kick contribution,
\begin{equation}
    P_\phi G P_\phi\neq 0,
\end{equation}
and the zeroth-order eigenstates hybridize at $O(1)$ across adjacent $P_{\rm dimer}$ sectors, lifting degeneracy. Then, the lifetime is short and the memory dephases within $O(1/\epsilon)$ cycles.

The main text discussed that the fidelity of the protected states collapses at the sideband resonances $r\omega=2|1-\Delta|$, Eq.~(22), where the kick-connected pair enters a common eigenphase multiplet. Figure~\ref{lifetime_vs_d} shows the lifetime $t^{*}$ as a function of $\Delta$ for the fully polarized initial state, with $L=10$ (circles) and $L=12$ (squares). Off resonance $t^{*}$ increases with $L$, consistent with Fig.~1. At the resonance points, $\Delta_r=1+r\omega/2$,  lifetime collapses. For our choice of $\omega=3$, $\Delta_r$ turns out to be $1,2.5,4,\cdots$. Secular mixing within the eigenphase multiplet happens between adjacent charge sectors for $r \ne 0$. The collapse at $\Delta=1$ is the special case of $r=0$ and is due to exact kick degeneracy, where $E_+=E_0$ and $[G,H(\lambda)]=0$.


\clearpage

\renewcommand{\thesection}{S\arabic{section}}
\renewcommand{\theequation}{S\arabic{equation}}
\setcounter{equation}{0}
\renewcommand{\thefigure}{S\arabic{figure}}
\setcounter{figure}{0}




\end{document}